\documentclass[aps,prl,twocolumn,groupedaddress,showpacs,
preprintnumbers,amsmath,amssymb]{revtex4}
\usepackage{graphicx}
\usepackage{dcolumn}
\usepackage{bm}
\usepackage{amsmath}

\begin{document}



\title{Kohn-Sham Density Functional Inspired Approach to 
Nuclear Binding}

\author{M. Baldo$^{1}$, P. Schuck$^{2,3}$ and X.
Vi\~nas$^4$}
\affiliation{
\it $^{1}$Instituto Nazionale di Fisica Nucleare, 
Sezione di 
Catania 
\centerline{Via Santa Sofia 64, I-95123 Catania, Italy} \\
\centerline{$^2$ Institut de Physique Nucl\'eaire, CNRS, UMR8608, 
  Orsay, F-91406, France} \\
\centerline{$^3$Universit\'e Paris-Sud, Orsay, F-91505, France}
\centerline{$^4$Departament d'Estructura i Constituents 
de la Mat\`eria,} \\
\centerline{ Universitat de Barcelona,
 Av. Diagonal 647, E-08028 Barcelona, Spain}}



\begin{abstract}

A non-relativisitic nuclear density functional theory is 
constructed, not as done most of the time, from an
effective density dependent nucleon-nucleon force but  
 directly introducing in the 
functional results from microscopic nuclear  
and neutron matter Bruckner G-matrix calculations at 
various densities.
A purely phenomenological finite range part to account for surface properties
is added. The striking result is that only 
four to five adjustable parameters, spin-orbit included, 
suffice 
to
reproduce nuclear binding energies and radii with the same 
quality as obtained with the most performant effective forces.
In this pilot work, for the pairing correlations, simply a density 
dependent zero range force is adopted from the 
literature. Possible
future extensions of this approach are pointed out.

\end{abstract}

\maketitle

It is common use in nuclear physics to describe
properties of nuclei in
mean-field approximation, using effective density
dependent
nucleon-nucleon interactions. Prototypes of these
interactions are the
so-called Skyrme-forces \cite{vauth} which are of 
zero range and density dependent. There also exist finite 
range versions, like
the Gogny force \cite{gog}
and the relativistic mean field (RMF) approach 
\cite{Ring}. The number of parameters which enter these
effective forces
is typically around ten and they are adjusted 
to reproduce
finite nuclei and 
some nuclear matter properties. 
However,
recently also data from a theoretically determined neutron matter 
Equation of State (EOS) \cite{sly4} 
have been used as input (see also an older attempt in this 
direction in \cite{Lamb}).
Generally these forces give rise to an
effective nucleon mass $m^*<m$ with typically
$m^*/m \simeq 0.7$ in the non-relativistic 
framework. With these
ingredients nuclear mean field theories are very
successfull to describe
nuclear properties as, e.g. binding energies, radii, but
also excited
states, fission barriers, and many things more 
\cite{Li,Ring}. These
effective forces are
still under debate and constantly refined to account for
the ever
increasing set of nuclear data \cite{sly4,brown98}. For 
instance the
evolution with isospin, to be considered for nuclei approaching the
neutron drip
 as involved in stellar
nucleo-synthesis, is
an active line of research. Also, usually ( apart from the Gogny force),
additional parameters are needed to characterise nuclear pairing.


The main objective of the present study is to show 
that within  an approach inspired by the Kohn-Sham Density Functional Theory 
(KS-DFT), like 
it is presently
still used extensively in, e.g., condensed matter, 
chemistry, and atomic physics 
\cite{jones,eschrig,Mat02} ( see, however, also \cite{Per01,Per03,Per05} 
for 
updated 
versions 
of the 
universal exchange-correlation energy including gradients and kinetic energy 
density 
dependences), a very efficient energy density functional 
for finite nuclei can be built up  using explicitly 
a fully microscopic 
input from nuclear and neutron matter calculations. We are, however, aware of 
the fact that the applicability of the KS-DFT approach to self-bound systems, 
as nuclei, is not obvious and that presently vivid debate about this topic 
is going on \cite{Krei01,Eng07,Dob07,Gir07,workshop}. In 
this work we shall not 
contribute 
anything 
fundamental to this discussion but rather adopt a pragmatic point of view: 
try and see! The KS-DFT scheme is somewhat different from the usual Skyrme 
or Gogny approaches \cite{vauth,gog} where, as mentioned, most of 
the time an effective {\it 
force} is constructed. However, several authors also interpret the Skyrme 
force as based on DFT \cite{Brack,Rein95} and a few authors have, in 
the 
past, constructed local Energy Density Functionals (EDF's) which cannot be 
related to underlying effective forces \cite{Lomb73,Fay00,Bulgac}. 
See also \cite{BHR} for a review of different mean field calculations used in 
nuclear physics.
In detail 
, however, all these nuclear approaches differ from what is practice in 
condensed matter physics (see below). The basis of KS-DFT lies in the 
Hohenberg-Kohn (HK) theorem 
 \cite{HK}, which states that for a Fermi system, with a non-degenerate 
ground state, the total energy can be expressed as a functional of the density 
$\rho({\bf r})$ only. Such a functional reaches its variational minimum
when evaluated with the exact ground state density.  Furthermore, for
practical reasons, in the KS-DFT method one introduces an auxiliary
set of $A$ orthonormal single particle wave functions $\psi_i({\bf r} )$, 
where $A$ is the number of particles, and the density is assumed to be given 
by
\begin{equation}
 \rho( {\bf r} ) \, =\, \Sigma_{i,s,t} | \psi_i( {\bf r},s,t ) |^2
\label{eq1} \end{equation}
\noindent where $s$ and $t$ stand for spin and iso-spin indices. The 
variational procedure to minimise the functional is performed in
terms of the orbitals
instead of the density. Usually in condensed matter and 
atomic physics the HK functional $E[\rho({\bf r})$] is 
split into two parts: $E = T_0[\rho]+W[\rho]$ \cite{KS}. 
The first piece $T_0$ corresponds to  the uncorrelated 
part of the kinetic energy and within the KS method it 
is written as
\begin{equation}
 T_0=\frac{{\hbar}^2}{2m} \sum_{i,s,t} \int 
d^3r|\nabla \psi_i( {\bf r},s,t ) |^2.
\label{eq2} \end{equation}
\noindent
The other piece $W[\rho]$ contains the potential energy 
as well as the correlated part of the kinetic energy.

 Then, upon variation, one gets a closed
set of $A$ Hartree-like equations with an effective
potential, the functional derivative of $W[\rho]$ with 
respect to the local
density $\rho({\bf r})$. Since the latter depends on the density, and
therefore on the
 $\psi_i$'s, a self-consistent procedure is 
necessary. Still the equations are
exact but they only can be of some use if a reliable approximation is found
for the otherwise unknown density functional $W[\rho]$. 
In the KS-DFT formalism the exact ground
state wave function is actually not known, the 
density being the basic quantity. 

In nuclear physics, contrary to the situation in 
condensed matter and atomic physics, the 
contribution of the spin-orbit interaction to the energy 
functional is very 
important. 
 Non-local contributions have been included in DFT in several ways already 
long ago 
(see \cite{Eng03} for a recent review of this topic). Consequently,
the spin-orbit 
part also can be split in an uncorrelated part 
$E^{s.o.}$ plus a remainder. The form of the uncorrelated 
spin-orbit part is taken exactly as in the 
Skyrme \cite{vauth} or Gogny forces \cite{gog}. 




We thus write
for the functional in the nuclear case
$E = T_0 + E^{s.o.} + E_{int} + E_C$,
where we explicitly split off the Coulomb energy $E_C$ 
because it is a quite distinct part in the Hamiltonian.
It shall be treated, as usual, at lowest order, i.e. the 
direct term plus the exchange contribution in the Slater 
approximation, that is
$E_C^H=
(1/2) \int \int d^3rd^3r' \rho_p({\bf r})|{\bf r}-{\bf 
r'}|^{-1}
\rho_p({\bf r'})$,
 and 
$E_C^{ex} = -(3/4)(3/\pi)^{1/3} \int d^3r {\rho_p({\bf 
r})}^{4/3}$
with $E_C=E_C^{H} + E_C^{ex}$ and $\rho_{p/n}$ the 
proton/neutron density.
\par Let us now discuss the nuclear energy functional
contribution $E_ {int}[\rho_n,\rho_p]$ which contains 
the nuclear potential energy as well as 
additional correlations. 
We shall split it in a finite range term 
$E_ {int}^{FR}[\rho_n,\rho_p]$ to account for correct 
surface properties and a bulk correlation part 
$E_ {int}^{\infty}[\rho_n,\rho_p]$
that we take from a microscopic infinite 
nuclear matter calculation \cite{BMSV} as we will 
discuss below. Thus our final 
KS-DFT -like functional reads:
\begin{equation}
E = T_0 + E^{s.o.} + E_{int}^{\infty} + E_{int}^{FR} 
+ 
E_C.
\label{eq6} \end{equation}

For the finite range term we make the 
simplest 
phenomenological ansatz possible
\begin{eqnarray}
E_{int}^{FR}[\rho_n,\rho_p ] &=& 
\frac{1}{2}\sum_{t,t'}\int
\int d^3r d^3r'\rho_{t}({\bf r})
v_{t,t'}({\bf r}-{\bf r'})\rho_{t'}({\bf r'} )
\nonumber \\
&-& \frac{1}{2}\sum_{t,t'}
\gamma_{t,t'}\int
d^3r {\rho_{t}({\bf r})} \rho_{t'}({\bf r})
\label{eq7} \end{eqnarray}
with $t=$ proton/neutron and $\gamma_{t,t'}$ the volume integral of
$v_{t,t'}(r)$.
The substraction in (\ref{eq7}) is made in order not to 
contaminate the bulk
part, determined from the microscopic infinite 
matter calculation. Finite range terms have 
already 
been used earlier, generalizing usual Skyrme functionals (see e.g 
\cite{BKN,Umar,Fay00}). In this study, for 
the finite range form factor $v_{t,t'}(r)$ we make a simple Gaussian ansatz:
$v_{t,t'}(r)=V_{t,t'}e^{-r^2/{r_0}^2}$. We choose a 
minimum of three open
parameters: $V_{p,p}=V_{n,n}=V_L, V_{n,p}=V_{p,n}=V_U$, and $r_0$. 

The only undetermined and most important piece in 
(\ref{eq6}) is then the bulk 
contribution $E_{int}^{\infty}$. 
In condensed matter, chemistry and atomic physics, this quantity implies 
the EOS of interacting electrons and for the KS-DFT scheme, most of the time, 
it is obtained from Quantum Monte-Carlo (QMC) calculations, the results of 
which are then accurately represented by a fit function (not necessarily a 
polynomial, as we use later). As already mentioned, we obtain 
$E_{int}^{\infty}$ from 
microscopic infinite matter calculations, using a realistic bare force, 
together with a converged hole-line expansion 
\cite{BMSV}.
We first reproduce by
interpolating functions the {\it correlation} part of the 
ground 
state energy per particle of 
symmetric and pure neutron matters,
and then make a quadratic interpolation for asymmetric 
matter.
Finally the total correlation contribution to the energy 
functional in local density approximation reads:
\begin{equation}
E_{int}^{\infty}[\rho_p, \rho_n] =  
\int d^3r \big[ P_s(\rho) (1 - 
\beta^2) + P_n (\rho)\beta^2 \big] \rho  
\label{eq4a}
\end{equation}
 where $P_s$ and $P_n$ are two interpolating polynomials 
for symmetric and
pure neutron matter, respectively, at the density
$\rho = \rho_p + \rho_n$, and
$\beta = (\rho_n - \rho_p)/\rho$ is the asymmetry 
parameter.
For $P_s$ we took
( $x = \rho / \rho_0$, with $\rho_0$=0.17 fm$^{-3}$, 
see below )
\begin{equation}
 P_s( \rho ) =  \left\{ \begin{array}{lr}
                 \sum_{k=1}^5 b_k^{(s)} x^k  &   x< 1  \\
                                 &  \\
                 P_s(\rho_0) + a_1\cdot (x-1) + a_2\cdot 
(x-1)
^2  &  x > 1
                 \end{array} \right.
\label{sym}
\end{equation}
where the coefficients ( energy/particle in MeV) are 
given in
Table I.
The two forms match at $x = 1$ ($\rho = \rho_0$) up to 
the second derivative.
This functional form  can be used up to $\rho = 
0.24$ fm$
^{-3}$, which is
the interval where the independent fit of 
the microscopic calculation has been performed.
\par
A similar expression holds for $P_n$,
\begin{equation}
 P_n( \rho ) = \sum_{k=1}^5 b_k^{(n)} x^k
\label{neu}
\end{equation}
where again the coefficients are displayed in Table I, 
which
is valid in the same density interval. The interpolating
polynomial for
symmetric matter has been constrained to allow a 
minimum
exactly at the energy
$E/A$ = - 16 MeV and Fermi momentum $k_F$ = 1.36 fm$^{-1}$, i.e. $\rho_0$=
0.17 fm$^{-3}$.
This is within the
uncertainty of the numerical microscopic calculations of 
the
EOS \cite{fnote1}.
The constrained fit was performed by keeping the EOS as smooth as possible, 
thus allowing for some very small deviations from the microscopic calculations 
below saturation density. An interpolating fit which goes exactly through the 
calculated EOS, as performed in \cite{BMSV}, gives a not good enough 
saturation point (typically E/A = -15.6 MeV, $k_F$=1.38 fm$^{-1}$). In 
this context 
it should be noticed that the microscopic calculations of Ref.\cite{Akm98} 
have been modified, in order to get a good saturation point, by an ad hoc 
correction which amounts to more than 2 MeV in energy ( see, e.g. \cite{Mar07} 
 for a presentation of the uncorrected EOS of \cite{Akm98}). As 
discussed in \cite{Mar07}, the low density behavior of the nuclear 
matter EOS is quite intricate and usually not reproduced by Skyrme and Gogny 
functionals ( see also ref. \cite{BMSV}), missing quite a substantial part 
of binding. We show our EOS for nuclear and neutron matter in Fig. 1. One 
sees that for all densities relevant to finite nuclei the EOS is very 
accurately reproduced by the interpolation (continuous lines). The bulk 
part $E_{int}^{\infty}$ of our functional, directly related to bare NN 
and NNN forces, is, therefore, determined once and for all and we will use 
it in (3) together with LDA. 
\begin{figure}
\includegraphics[width=7.5cm]{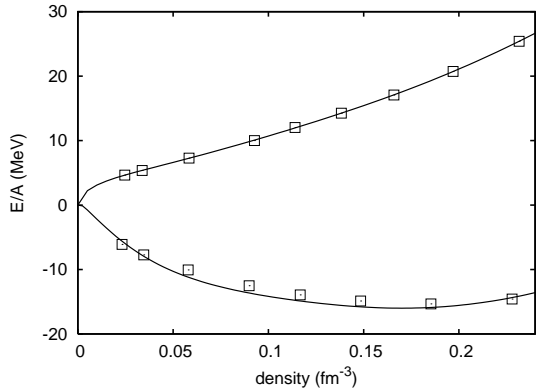}
\caption{\label{Figure1} EOS of symmetric and neutron 
matter obtained by the microscopic calculation 
(squares) and the corresponding polynomial fits (solid 
lines).} \end{figure}
  
The only open parameters are, therefore, the 
ones contained 
in the finite range surface part, Eq.(\ref{eq7}), and the strength of the 
spin-orbit 
contribution. We, thus, follow {\it exactly} the strategy employed in the 
above mentioned 
interacting electron systems.  
On the contrary, up to now, in nuclear physics, 
almost exclusively
a different strategy was in use (see, however, Ref.\cite{NV73} for a 
somewhat similar approach to our): functionals like the one of (\ref{eq6}), 
i.e. bulk, 
surface, etc., were 
globally parametrized with typically of the order of ten parameters which, 
then, were 
determined fitting {\it simultaneously} some equilibrium nuclear matter 
(binding energy 
per particle, 
saturation density, incompressibiliy, etc.) and finite nuclei properties. 
However, in this way, bulk and surface are not 
properly separated and
early attempts used to miss important infinite nuclear matter properties, as, 
e.g. 
stability of neutron matter at high density \cite{gog} and other stability 
criteria. Modern Skyrme 
forces , like the
Saclay-Lyon (SLy) ones, explicitly use the high density part 
($\rho/\rho_0 > 0.65$) of 
microscopic neutron matter
calculations for the EOS in the fitting procedure \cite{sly4} and thus avoid 
collapse.
Therefore, modern functionals usually reproduce reasonably well 
microscopically determined 
EOS for neutron and nuclear matter \cite{BMSV}. Examples are, among others, 
 the 
SLy-forces  
\cite{sly4} (see Fig.1 in \cite {BMSV})
and the Fayans functional (see e.g. Fig. 3 in  \cite{Fay00}). 
It is thus evident that the 
procedure adopted in this work is different on a qualitative level from 
the usual and allows, via the fit, to reproduce very accurately the 
microscopic 
infinite matter results in the whole range of densities considered.  
This may be important for surface properties and neutron skins in 
exotic nuclei, what shall be investigated in the future.
Let us also mention, in this context that for the electron systems 
mentioned before, the DFT practitioners demand to have extremly precise
EOS as LDA input to their calculation.                    
         
For open shell nuclei, we still have to add  pairing. 
The formal generalization of the rigorous HK theorem to paired system 
has been given
in Ref. \cite{Oli88}.
In  the present  work our
main objective is to set up the KS-DFT scheme for the non 
pairing part, thus we add
pairing in a very simple way within the BCS approach. 
For this we simply take the
density dependent delta force defined in Ref \cite{GSGS} for 
$m=m^*$  with the same parameters and in particular with the same cutoff. 
As far as this amounts to a cutoff of $\sim$ 10 MeV into the 
continuum for finite nuclei, we have to deal with 
single-particle energy levels lying in the continuum. We have simulated it by 
taking 
in the pairing window all the quasi-bound levels, i.e. the levels retainded 
by the 
centrifugal (neutrons) and centrifugal plus Coulomb (protons) barriers. This 
tratment of the continuum works properly, at least for nuclei not far from 
the stability valley as it has been extensively shown in \cite{Estal}. 
A more 
sophysticated procedure, as, e.g, the one proposed by Bulgac 
et al \cite{Bulgac}, may turn out to lead to better results. Since we do not
want to concentrate on pairing here, this shall be investigated in forthcoming 
work.

In our calculations the two-body center of mass 
correction has been included in the self-consistent calculation using 
the pocket formula, 
based on the harmonic oscillator, derived in 
Ref.\cite{BSM} which nicely reproduces the exact 
correction as it has been shown in \cite{STV}. Our functional is now fully 
defined and, henceforth, we call it BCP-functional.

In this exploratory investigation, we
fitted two sets of parameters. We have considered only spherical nuclei which
we chose as given below. The two fits were obtained in i) optimising ground
state energies only, ii) optimising ground state energies together with 
the charge rms radii. The open parameters, $V_L$, $V_U$, $r_0$ of 
Eq.(\ref{eq7}) as well as the spin-orbit strength $W_0$, 
are fitted to reproduce the binding
energies of the nuclei $^{16}$O, $^{40}$Ca, $^{48}$Ca, 
$^{56}$Ni, 
$^{78}$Ni,
$^{90}$Zr, $^{116}$Sn, $^{124}$Sn, $^{132}$Sn, $^{208}$Pb 
and
$^{214}$Pb in the case of the parameter set BCP1 and
additionally the charge rms
radii ($r_c = \sqrt{r_p^2 +0.64}$ fm) of the nuclei 
$^{16}$O, 
$^{40}$Ca,
$^{48}$Ca,$^{90}$Zr, $^{116}$Sn, $^{124}$Sn, $^{208}$Pb 
and $^{214}$Pb
for the parameter set BCP2. Experimental binding energies
and charge radii are taken from References \cite{AW} and
\cite{Angeli} respectively. In table II we give the 
obtained parameter sets of fits (i) and (ii).

Our intention is to compare our results from the 
BCP-functionals with the ones obtained from some of the most
performant effective nucleon-nucleon forces available, 
namely D1S \cite{D1S}, NL3 \cite{NL3} and SLy4 \cite{sly4}. 
To this end, we have calculated the binding 
energies and charge radii of 161 even-even spherical nuclei 
(in line with the NL3 calculations reported in 
\cite{NL3res}). 
In Table III we report the energy and charge 
radii rms deviations between the 
corresponding experimental values and the theoretical 
ones obtained using the D1S force \cite{D1Sres}, the NL3 
parametrization \cite{NL3res}, the Skyrme interaction SLy4 
\cite{sly4} and our BCP1 and BCP2 
functionals. 
We also display in Figs.2-3 the differences between the 
theoretical and experimental energies and charge radii in 
the range between $^{16}$Ne and $^{224}$U 
calculated with 
the BCP1 functional in comparison with the same quantities 
obtained from the D1S effective force \cite{D1Sres}.

In Table IV we display the two-neutron separation energies for
some magic nuclei predicted by the BCP1 and BCP2 functionals as 
compared with the same results provided by D1S and NL3 as well as the 
corresponding experimental values.

We see that our theory
is at least as performant as the
other ones. 
In Table V we give the common bulk 
properties of both BCP-functionals. Incompressibility 
modulus $K_{\infty}$ and 
symmetry energy $J$ plus its derivative $L$ have quite 
acceptable values. For the average gap in $^{116}$Sn 
we obtain 1.5 MeV. For the drip line nucleus $^{176}$Sn
we get $E/A$=-6.57 MeV compared to $E/A$=-6.44 MeV (D1S)
and $E/A$=-6.67 MeV (NL3).

In a recent analysis, Bertsch et al.\cite{bertsch} 
have pointed out that also in Skyrme forces implicitly 
only four or five parameters are relevant. 
For example, in the SLy forces \cite{sly4}, six out of 
the twelve 
parameters are strongly constrained by nuclear and neutron matter properties.
However, the philosophy in the present work is different and, as we said 
before, is 
analogous to what is done in condensed matter physics. That is we use as an 
input
an analytic representation of a very accurate ab initio calculation of the 
nuclear/neutron EOS in the whole relevant range of densities, feed this 
into the functional via LDA and fit the remaining four parameters, which 
characterize the surface and spin-orbit contributions to the energy density, 
to finite 
nuclei properties. 
It is remarkable and, to 
some extend unexpected, that this scheme works indeed very well. It seems to 
give some credibility to the applicability of the KS-DFT scheme, 
even for self bound systems, as treated here (see also a discussion of 
this point in \cite{Gir07}).

\begin{table}[ht]
\caption{\label{Table1} Coefficients of the polynomial 
fits $P_s$, Eq.
(\ref{sym}) and $P_n$, Eq. (\ref{neu}) to the EOS of  
symmetric and 
neutron matters.}
\begin{center}
\begin{tabular}{rrrr}
k &  $b_k^{(s)}(MeV)$  &  $b_k^{(n)}(MeV)$ & $a_k$(MeV) \\
\hline
1  & -105.640069 & -43.985736 & -15.3563461  \\
2  &  167.700968 &  49.784439 &  16.4197441  \\
3  & -181.762432 & -42.400650 &  0.0  \\
4  &  103.166047 &  21.894382 &  0.0 \\
5  & -22.4990207 & -4.3071179 &  0.0  \\
\hline
\end{tabular}
\end{center}
\end{table}
\begin{table}[ht]
\caption{\label{Table2} Parameters of the Gaussian form factors and spin-orbit
 strength.}
\begin{center}
\begin{tabular}{rrrrrr}
& $r_0$(fm) & $V_L$ (MeV) & $V_U$ (MeV) & $W_0$ (MeV) \\
\hline
BCP1 & 1.05 & -93.520 & -60.577 & 113.829  \\
BCP2 & 1.25 & -33.700 & -32.483 & 110.812  \\
\hline
\end{tabular}
\end{center}
\end{table}
\begin{table}[ht]
\caption{\label{Table3}  Energies (in MeV) and charge radii 
(in fm) rms deviations. The numerical values of energies 
and charge radii calculated with the BCP1 and BCP2 
functionals used to obtain the rms values of this Table
are given in \cite{server}}
\begin{center}
\begin{tabular}{rrrrrr}
& BCP1 & BCP2 & D1S & NL3 & SLy4 \\ 
\hline
rms$_E$ & 1.775 & 2.057 & 2.414 & 3.582 & 1.711\\
rms$_R$ & 0.031 & 0.028 & 0.020 & 0.020 & 0.024\\
\hline
\end{tabular}
\end{center}
\end{table}

\begin{table}
\caption{\label{tab1}
Two-neutron separation energies $S_{2n}$ (in MeV) of some magic 
nuclei computed with BCP energy functionals, D1S force 
and NL3 parametrization 
Experimental values \cite{AW} are also displayed.}
\begin{ruledtabular}
\begin{tabular}{llccccc}
& & $^{16}$O  &  $^{40}$Ca &  $^{48}$Ca 
   & $^{132}$Sn & $^{208}$Pb \\
\hline
& BCP1 &
25.57 & 26.62 & 15.69 & 11.94 & 14.77 \\
& BCP2 &
25.82 & 26.93 & 15.97 & 12.05 & 14.79  \\
& D1S &
29.57 & 29.60 & 15.68 & 12.79 & 14.89 \\
& NL3 &
28.41 & 29.72 & 16.36 & 12.55 & 13.99 \\
& exp &
28.87 & 28.93 & 17.22 & 12.56 & 14.11 \\

\end{tabular}
\end{ruledtabular}
\end{table}

\begin{table}[ht]
\caption{\label{Table4.} Common infinite nuclear matter
properties of both BCP functionals}
\begin{center}
\begin{tabular}{rrrrrrr}
& $B/A$(MeV) & $\rho_0$ (fm$^{-3}$) & $m/m^*$ & $J$ (MeV) &
$L$(MeV) & $K_{\infty}$(MeV) \\
\hline
& -16.00 & 0.17 & 1.00 & 33.55 & 56.39 & 249. \\
\hline
\end{tabular}
\end{center}
\end{table}
\begin{figure}
\includegraphics[width=7.5cm,angle=-90]{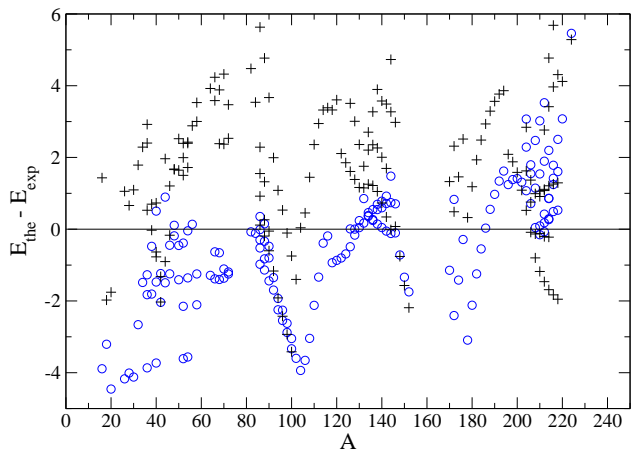}
\caption{\label{Figure2} Differences between the 
theoretical and
experimental energies. Calculations are 
performed
using the BCP1 functional (open circles) and the D1S 
parameter set (crosses) \cite{D1Sres}.}
\end{figure}
\begin{figure}
\includegraphics[width=7.5cm,angle=-90]{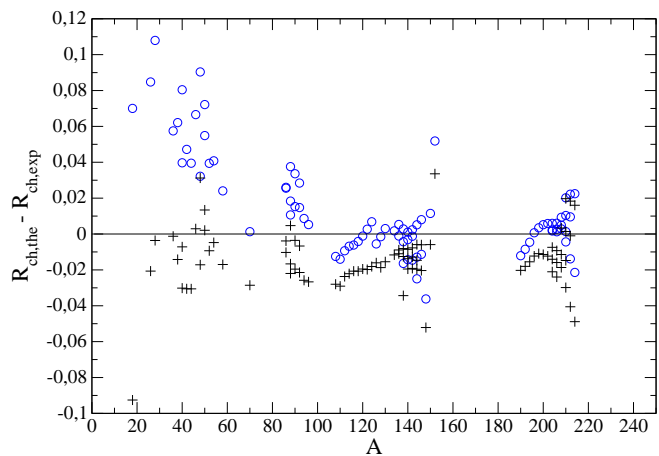}
\caption{\label{Figure3}
 Differences between the theoretical and
experimental charge rms radii. Calculations are performed
using the BCP1 energy functional (open circles) and D1S 
parameter set (crosses) \cite{D1Sres}.}
\end{figure}

In conclusion, we have succesfully applied a 
Kohn-Sham 
Density Functional Theory inspired approach to ground state properties of 
nuclei, 
quite in analogy to what is usually done in e.g. condensed matter physics and 
for atoms and/or molecules
\cite{jones,eschrig}, however, at variance with the procedure hitherto 
almost exclusively used 
in nuclear physics. This inspite of the fact that the applicability  of 
KS-DFT for 
self-bound systems is, so far, an unsettled problem. The bulk part of the 
functional is 
given for all relevant densities and asymmetries, once and for all, from microscopic 
results using the converged hole-line expansion 
based on  
realistic bare forces \cite{BMSV}. To  this a phenomenological surface part 
with three adjustable parameters is added. A fourth parameter is the strength 
of the spin-orbit interaction for which we get values close to the usual 
ones in Gogny and Skyrme forces. The microscopic  nuclear matter EOS has been 
fine tuned to pass through $E/A$ = -16.00 MeV at saturation. This may be 
considered as the fifth adjusted parameter. In this pilot study we took 
for the pairing part of the functional the simplest possible 
procedure and adopted a previously adjusted density 
dependent $\delta$-force \cite{GSGS}. 
 The fit of open
parameters was done only for spherical nuclei. In view of the surprisingly 
small number of adjustable parameters, the results are very encouraging and 
well compete with the most performant mean field theories presently in use. 
In addition we have preliminary results which show that  
our BCP-functional also yields excellent results in the deformed case,
again in very close agreement with those obtained with the Gogny D1S force 
\cite{Rob08}. Our study thus seems to support the 
applicability of the KS-DFT scheme also of self-bound systems.
A  more elaborate parameter search 
with different flexibilites 
than in (\ref{eq7}) is an important task for the near 
future.

We are indebted to C.Maieron for valuable discussions. 
One of us (P.S.) is greatful to M.Casida for useful 
informations. This work is partially supported by 
IN2P3-CICYT and INFN-CICYT. X.V. acknowledges Grant 
Nos.FIS2005-03142 (MEC, Spain and FEDER) and 
2005SGR-00343 (Generalitat de Catalunya).



\begin{thebibliography}{99}

\bibitem{vauth} D. Vautherin, D.M. Brink,  
 Phys. Rev. {\bf C5} (1972) 626.

\bibitem{gog} J. Decharg\'e, D. Gogny, Phys. {\bf C21} (1980) 1568.

\bibitem{Ring} P. Ring, Prog. Part. Nucl. Phys. {\bf 37} 
(1996) and 
references therein.

\bibitem{sly4} E. Chabanat, P. Bonche, P. Haensel, J.
 Mayer and R. Schaeffer,
Nucl. Phys. {\bf A627} (1997) 710; {\bf A635} (1998) 231.


\bibitem{Lamb} D. Q. Lamb, J. M. Lattimer, C. J. Pethick, D. G. 
Ravenhall, Nucl. Phys. {\bf A360}, 459 (1984).

\bibitem{Li} Li Guo-Quiang, 
J. of Phys. {\bf G17} (1991) 1.

\bibitem{brown98} B.Alex Brown,
 Phys. Rev. {\bf 58} (1998) 220.

\bibitem{KS}
W. Kohn and L. J. Sham, Phys. Rev. {\bf 140},
A1133 (1965).

\bibitem{jones}
R. O. Jones and O. Gunnarsson, Rev. Mod. Phys. {\bf 61},
689 (1989).

\bibitem{eschrig}
H. Eschrig,
The Fundamentals of Density Functional Theory
(B. G. Teubner, Stuttgart, 1996).

\bibitem{Mat02}
A. E. Mattsson, Science {\bf 298}, 759 (2002).

\bibitem{Per01}
J. P. Perdew and K.Schmidt,
Density Functional Theory and Its Applications to Material, V. Van Doren 
et al. Eds. CP577 (American Institute of Physicas, Melville NY, 2001.

\bibitem{Per03}
Jianmin Tao, J. P. Perdew, V. N. Staroverov and G.
F. Scuseria, Phys. Rev. Lett. {\bf 91}, 146401 (2003).

\bibitem{Per05}
Jianmin Tao and  J. P. Perdew, Phys. Rev. Lett.{\bf 95}, 196403 (2005).

\bibitem{Krei01}
T. Kreibich and E. K. U. Gross, Phys. Rev. Lett. {\bf 86}, 2984 (2001).

\bibitem{Eng07} J. Engel, Phys. Rev. {\bf C75}, 014306 (2007).

\bibitem{Dob07} J. Dobackzewski, M. V. Stoitsov, W. Nazarewiczand P.-G. 
Reinhard, Phys. Rev. {\bf C76}, 054315 (2007).

\bibitem{Gir07} B. G. Giraud, B. K. Jenning ands B. R. Barret, 
arXiv:0707.3099 (2007); B. G. Giraud, arXiv:0707.3901 (2007)

\bibitem{workshop} First "FIDIPRO-JSPS Workshop on Energy 
Density Functionals in Nuclei", Jyvaskyla, October 25-27, 
2007; "Formal aspects of Nuclear Energy Density 
Functional Methods", Scalay, November 5-7, 2007. 

\bibitem{Brack}
M. Brack, Helvetia Physica Acta {\bf 58}, 715 (1985).

\bibitem{Rein95}
P.-G. Reinhard, and H. Flocard, Nucl. Phys. {\bf A584}, 467 (1995)
and references therein.

\bibitem{Lomb73}
R. J. Lombard, Ann. of  Phys. (NY) {\bf 77}, 380 (1973).

\bibitem{Fay00}
S. J. Fayans et al Nucl. Phys. {\bf 676}, 49 (2000) and references 
therein.

\bibitem{Bulgac}
Y. Yu and A. Bulgac, Phys. Rev. Lett. {\bf 90}, 222501 (2003).

\bibitem{BHR}
M. Bender, P.-H. Heenen and P.-G. Reinhard,
Rev. fo Mod. Phys. {\bf 75}, 121 (2003).

\bibitem{HK}
P. Hohenberg and W. Kohn, Phys. Rev. {\bf 136},
B864 (1964).


\bibitem{Eng03}
J. Engel,
Lecture Notes in Physics {\bf 620}, 56 (2003), C. Fiolhais, F. Nogueira 
and M. Marques Eds. (Springer-Verlag, Berlin, 2003).

\bibitem{fnote1}
The value $E/A=-16.00 MeV$ has nothing special and we did 
vary this parameter in slight proportions. Decreasing 
this value up to $E/A=-16.15 MeV$ still moderately 
improves the rms values for energies and radii (see 
Table III) but at the expense of an unacceptable high 
value of 
the incompressibility ($K_{\infty}$=279 MeV). We therefore took the above 
cited 
value as our optimal choice.

\bibitem{BMSV} M. Baldo, C. Maieron, P. Schuck and X.
Vi\~nas,
Nucl. Phys. {\bf A736}, (2004) 241 and references 
therein.

\bibitem{BKN} P. Bonche, S. Koonin and J. W. Negele,
Phys. Rev. {\bf C13}, 1226 (1976).

\bibitem{Umar}
A. S. Umar, M. R. Strayer, P.-G. Reinhard, K. T. R. Davies and J.-S. Lee,
Phys. Rev. {\bf C40}, 706 (1989). 

\bibitem{Akm98}
A. Akmal, V. R. Pandharipande and D. G. Ravenhall,
Phys. Rev. {\bf C58}, 1804 (1998).

\bibitem{Mar07}
J. Margueron, E. van Dalen and C. Fuchs, 
Phys. Rev. {\bf C76}, 034309 (2007).

\bibitem{NV73} J. W. Negele and D. Vautherin, 
Nucl. Phys. {\bf A207}, 298 (1973).

\bibitem{Oli88}
Oliverira, E. K. U. Gross and W. Kohn, Phys. Rev. Lett. {\bf 60}, 2430 
(1988).



\bibitem{GSGS}
E.~Garrido, P.~Sarriguren, E.~Moya de Guerra, and 
P.~Schuck,
Phys. Rev. C {\bf 60}, 064312 (1999)

\bibitem{Estal} M. Del Estal, M. Centelles, X. Vinas, Phys. Rev. {bf C}

\bibitem{BSM} M.N. Butler, D.W.L. Sprung and J. Martorell,
Nucl. Phys. {\bf A422}, 157 (1984).

\bibitem{STV} V.B. Soubbotin, V.I. Tselyaev and X.
Vi\~nas, Phys. Rev. {\bf C67} (2003) 014324.

\bibitem{AW} G. Audi and A.H. Waspra,
Nucl. Phys. {\bf A729}, 337 (2003); downloadable at 
www.csnsm.in2p3.fr/AMDC/ \\ 
masstables/Ame2003/mass.mas03.

\bibitem{Angeli} I. Angeli,
Atomic Data and Nuclear Data Tables {\bf 87}, 185 (2004).

\bibitem{D1S}
J.F. Berger, M. Girod and D.Gogny,
Comput. Phys. Commun. {\bf 63}, 365 (1991);
Nucl. Phys.
{\bf A502}, 85c (1989).

\bibitem{NL3} G.A. Lalazissis, J. K\"oning and P.
Ring,
Phys. Rev. {\bf C55}, 540 (1997).

\bibitem{NL3res} G.A. Lalazissis and S. Raman,
Atomic Data and Nuclear Data Tables {\bf 71}, 1 (1999).

\bibitem{D1Sres} http://www-phynu.cea.fr; S. Hilaire and 
B. Nerlo-Pomorska, private communication.

\bibitem{bertsch} G.F. Bertsch, B. Sabbey and M. 
Uusn\"akki,
Phys. Rev. {\bf C71}, 054311 (2005).

\bibitem{server} 
\begin{verbatim}
http://www.ecm.ub.es/~assum/taula_xv
\end{verbatim}

\bibitem{Rob08} 
L. M. Robledo, M. Baldo, P. Schuck and X. Vi\~nas,
work in progress.














\end{thebibliography}
\end{document}